\documentclass{agujournal2019}
\pdfoutput=1
\usepackage{url} 
\usepackage{lineno}
\usepackage{soul}
\graphicspath{{./figures/}}
\usepackage{amsmath}
\usepackage{lmodern}
\draftfalse

\journalname{Earth and Space Science}

\begin{document}

%
%


\title{Autonomous Detection of Particles and Tracks in Optical Images}

%
%




\authors{Andrew J.~Liounis\affil{1}, Jeffrey L.~Small\affil{2}, Jason C.~Swenson\affil{1}, Joshua R.~Lyzhoft\affil{1}, Benjamin W.~Ashman\affil{1}, Kenneth M.~Getzandanner\affil{1}, Michael C.~Moreau\affil{1}, Coralie D.~Adam\affil{3}, Jason M.~Leonard\affil{3}, Derek S.~Nelson\affil{3}, John Y.~Pelgrift\affil{3}, Brent J.~Bos\affil{1},  Steven R.~Chesley\affil{4}, Carl W.~Hergenrother\affil{5}, Dante S.~Lauretta\affil{5}}


\affiliation{1}{NASA Goddard Spaceflight Center, Greenbelt, MD 20771}
\affiliation{2}{The Aerospace Corporation, Chantilly, VA 20151}
\affiliation{3}{KinetX Space Flight Dynamics Practice, Simi Valley, CA 93065}
\affiliation{4}{Jet Propulsion Laboratory, California Institute of Technology, Pasadena, CA 91109}
\affiliation{5}{Lunar and Planetary Laboratory, University of Arizona, Tucson, AZ 85705}




\correspondingauthor{Andrew Liounis}{andrew.j.liounis@nasa.gov}




\begin{keypoints}
\item Autonomous techniques are described for the identification and tracking of particles in motion about a celestial body.
\item The techniques are demonstrated on images from the OSIRIS-REx mission to the asteroid (101955) Bennu.
\end{keypoints}

%
%

%
%


\begin{abstract}
    During its initial orbital phase in early 2019, the Origins, Spectral Interpretation, Resource Identification, and Security--Regolith Explorer (OSIRIS-REx) asteroid sample return mission detected small particles apparently emanating from the surface of the near-Earth asteroid (101955) Bennu in optical navigation images.
    Identification and characterization of the physical and dynamical properties of these objects became a mission priority in terms of both spacecraft safety and scientific investigation.
    Traditional techniques for particle identification and tracking typically rely on manual inspection and are often time-consuming.
    The large number of particles associated with the Bennu events and the mission criticality rendered manual inspection techniques infeasible for long-term operational support.
    In this work, we present techniques for autonomously detecting potential particles in monocular images and providing initial correspondences between observations in sequential images, as implemented for the OSIRIS-REx mission.
\end{abstract}

\section*{Plain Language Summary}
We present two algorithms for identifying and tracking numerous unknown objects in orbit about a celestial body without manual analysis.
These algorithms are applied by the OSIRIS-REx mission to track particles ejected from the active near-Earth asteroid Bennu. Results from the mission are shown.

\section{Introduction}
In January 2019, the Origins, Spectral Interpretation, Resource Identification, and Security--Regolith Explorer (OSIRIS-REx) spacecraft entered into orbit about the near-Earth asteroid (101955) Bennu.  
Shortly after entering orbit, ejection events were observed from the surface of Bennu that produced orbiting and hyperbolic particles, with some surviving up to multiple days \cite{lauretta}.
This observation triggered the decision to process images shortly following each downlink to (a) identify particle ejection events that could potentially impact spacecraft safety and (b) generate raw observables for downstream event reconstruction, particle characterization, and trajectory estimation.
These goals require timely processing of up to 150 images per day depending on the mission phase, as well as efficient cataloging and dissemination of the raw data.
This information is necessary both for spacecraft safety, to ensure that the spacecraft does not collide with any objects, and for the scientific and engineering insight that it provides about the asteroid environment \cite{leonard, pelgrift, lauretta}. 

Traditionally, identifying and tracking moving objects has been handled either by manual inspection of monocular images and other sensor measurements, typically through a process known as blinking, or through automated processes that look for objects moving with similar velocity in multiple frames \cite{blinking, modp1, modp2, spacewatch, neat}.
These techniques have largely been developed for terrestrial-based observations of planets, asteroids, and comets, where the distance between the observer and the target is very large, and thus the apparent motion in subsequent images is small and fairly linear.
The techniques also frequently rely on a human analyst directly interacting with the data either for extraction or verification.
While these methods are effective when there are a few bright objects that are captured in the field of view in each frame moving nearly linearly in the detector, they begin to break down when there is the potential for a large number of dim objects in each frame with substantial non-linear motion observed in the detector, as may be the case when encountering an active asteroid or comet. 

In this paper, we describe two algorithms to automate the identification and tracking of many dim, fast-moving objects in optical images.
The first algorithm autonomously extracts potential observations of objects from images using image processing techniques.
The results from this algorithm can be used to assist an analyst with a manual blinking process by highlighting all of the potential objects in each frame beyond what is seen in the raw image, or they can be fed to an automated linking and tracking algorithm.
The second algorithm autonomously generates potential linking and tracks of objects from frame to frame when fed with the sequential results from the first algorithm.
These algorithms were implemented using the Goddard Image Analysis and Navigation Tool (GIANT) \cite{giant} for monitoring particles during the OSIRIS-REx proximity operations at active asteroid Bennu.

The algorithms described in this paper provide a new capability for fast, autonomous detection and tracking of objects in orbit about a celestial body.
The autonomous detection technique is designed to enable detection of very dim objects in images where there is significant stray light due to an overexposed body located within the field of view.
The autonomous tracking technique is designed to link detections of the same object across numerous image frames when both the observer and the target are moving quickly with respect to each other over the time period of the observations, and it is also robust to spurious detections and multiple objects in the same frame.
Both of these techniques represent significant advancements in the current state of the art in space-based object tracking.

First, we describe the image processing steps required to extract potential objects from an optical image.
Then we discuss ways to filter these potential objects to reject those that are caused by noise only.
Next, we provide examples of how this data can be used to assist an analyst in the manual linking of objects from frame to frame.
Finally, we describe a multiple object tracking extended Kalman filter (EKF) autonomous linking and tracking algorithm that can detect potential orbits of observed objects.
Throughout the algorithm descriptions, we provide examples of their application to images taken by the NavCam 1 navigation camera onboard the OSIRIS-REx spacecraft \cite{tagcams}.

\section{Autonomous Detection of Potential Objects in Optical Images}
\begin{figure}[p!]
    \centering
    \includegraphics[width=0.75\textwidth,trim=0in 0in 0in 0in,clip]{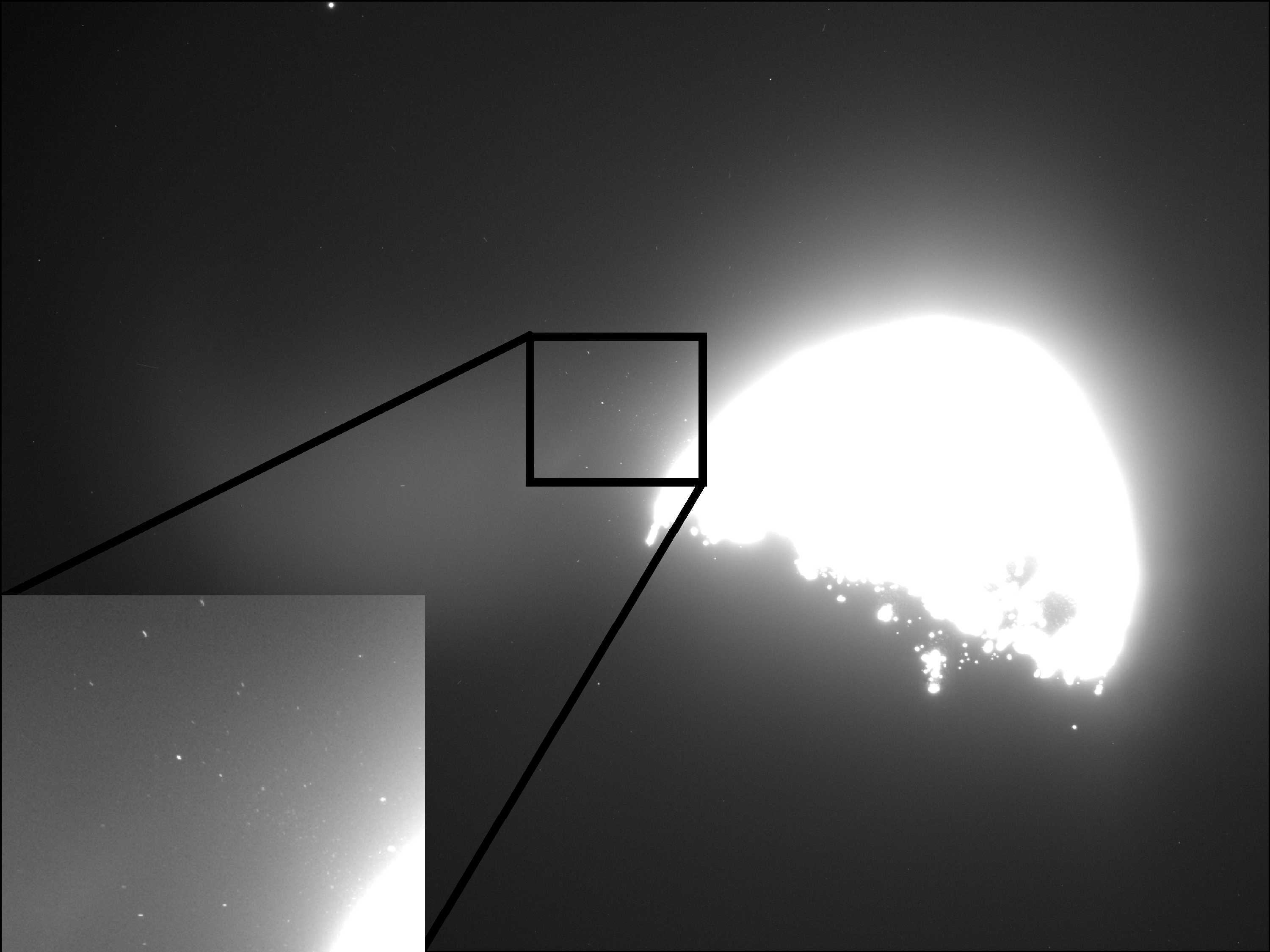}
    \caption{An example image containing particles and the overexposed extended body Bennu taken by the OSIRIS-REx NavCam 1 navigation camera on 6 January 2019.  The particles can be seen as the bright spots located to the immediate left of Bennu (and in the magnified call-out in the lower left).}
    \label{fig:example_unprocessed}
\end{figure}

Consider an image in which there are numerous small, dim objects in the field of view and one (or more) large, bright objects creating significant stray light throughout the detector.
An example of this is shown in Fig.~\ref{fig:example_unprocessed} taken on 6 January 2019 by the NavCam 1 navigation camera on the OSIRIS-REx spacecraft.
The objective is to extract all of the observations of the particles from the image in an autonomous manner.
If we assume that the majority of the objects are much smaller than the resolution of the detector in the angular sense, then we can assume that observations of each object are dominated by the point spread function (PSF) of the camera, and thus we need to identify bright spots that resemble a typical PSF.
Because this is similar to the lost-in-space problem for attitude determination using a star field, we can borrow many of the steps that are typically used to extract stars from an image to extract these particles.
Here we largely follow the steps in \citeA{astrometry.net} with some modifications needed for this specific application.

\begin{figure}[h!]
    \centering
    \includegraphics{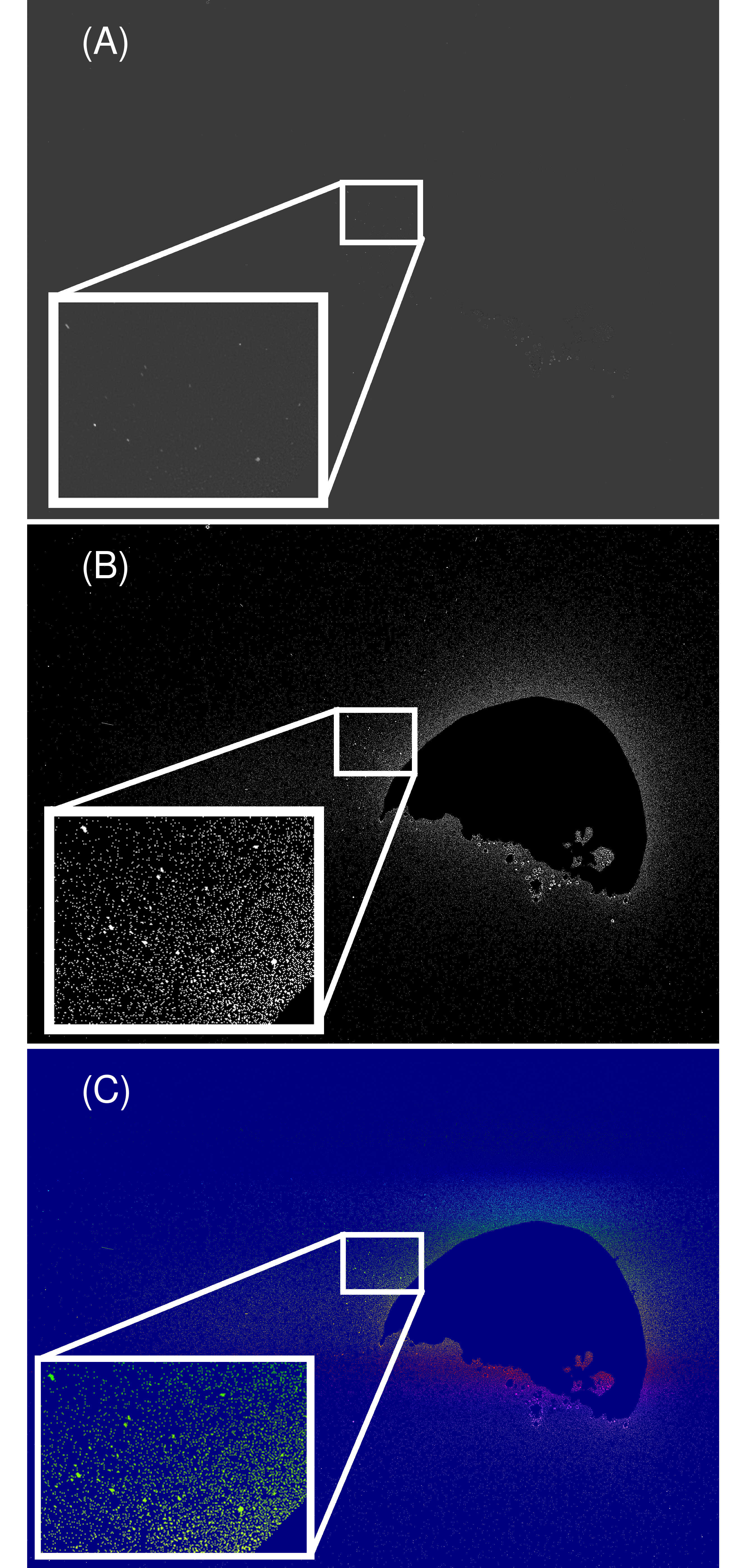}
    \caption{The image processing steps applied to identify potential particles in the images.  (A) The example image from Fig.~\ref{fig:example_unprocessed} after applying the flattening algorithm.  Most of the stray light due to the over-exposed asteroid has been removed and the small particles are more noticeable. (B) The flattened image from (A) after applying the thresholding algorithm.  The bright points in the image have been captured and assigned values of True (white) while the background has been ignored and assigned values of False (black). The terminator also results in many ``detections'' due to the rocky surface of Bennu. (C) The labeled image generated by applying a connected components algorithm to the binary image in (B).  Isolated blobs of bright pixels have been assigned the same label, and each unique blob shows as a different color. Each sub-figure includes a magnified call-out of the region of interest in the lower left.}
    \label{fig:example_image_processing1}
\end{figure}

We begin by flattening the image by subtracting a $5\times5$ median-filtered version of the image from the original image as is done in \citeA{astrometry.net}:
\begin{linenomath*}
\begin{equation}
    \mathbf{I}_{flat}=\mathbf{I} - \text{medfilt}\left(\mathbf{I}\right)
\end{equation}
\end{linenomath*}
where $\mathbf{I}$ is the original 2D grayscale image, $\text{medfilt}$ applies a 2D median filter, and $\mathbf{I}_{flat}$ is the flattened image.
This step removes some of the effects of stray light and elevated background signal and allows small, dim detections to stand out.
The result of this processing on the example image is shown in Fig.~\ref{fig:example_image_processing1} (A).

Now we need to determine the rough level of total noise (read noise, shot noise, etc.) in the intensity values (data number, DN) for each pixel for each image.  
This can be done in two different ways.
If the detector has covered active pixels (pixels that are active but are not exposed to an actual light source), as is common in many engineering and science instruments, then we can take the standard deviation of these covered pixels to estimate the noise level of the detector during each image.
If the detector does not have covered active pixels, then we can estimate the noise level by selecting a few thousand random pairs of pixels, computing the difference in each pair, and then computing the standard deviation of the differences, as is done in \citeA{astrometry.net}.  
If the latter technique is being used and there is a large overexposed body in the field of view, it is most useful to use an outlier rejection criteria on the DN differences to throw out pairs where one pixel is on the body and one is off the body and avoid an overly high noise estimate.
The modified z-score \cite{z-score} is useful for this task.

Given the rough noise level in the image, we can threshold the median-flattened image to identify all pixels that are some sigma above the background:
\begin{linenomath*}
\begin{equation}
    \mathbf{I}_{bin} = \mathbf{I}_{flat} \geq x\sigma
\end{equation}
\end{linenomath*}
where $\mathbf{I}_{bin}$ is the binary thresholded image, $\sigma$ is the noise level for the image computed in the previous step, and $x$ is the sigma multiplier.
The pixels where this criterion is met can now be said to be ``interesting pixels'' that require further investigation.
The number of sigma above the noise level that is used to perform the threshold can be adjusted higher or lower, depending on how dim detections are expected to be and how tolerant of false detections future processes are, but in our experience, as in \citeA{astrometry.net}, a value of 8 has proven to be effective.
The result of the noise estimation and thresholding on the example image is shown in Fig.~\ref{fig:example_image_processing1} (B).

With the binary thresholded image available, we now need to stitch together blobs of interesting pixels into single observations.
This can be done using a standard 8-connectivity connected components algorithm \cite{connected}.
Labeling the image in this way enables us to consider individual interesting blobs, instead of individual pixels.
This is especially helpful because we would typically expect the brightest objects to have signal in multiple pixels that meet the threshold requirements, and we want to consider these interesting pixels together.
The result of the connected components labeling on the example image is shown in Fig.~\ref{fig:example_image_processing1} (C).

We can now process each blob to find the center of the brightness representing each observation.
There are a number of different ways this can be performed, with varying levels of accuracy.
These range from simple algorithms, such as picking the peak pixel for each blob, to more complicated but more accurate algorithms, such as fitting a representative PSF to the DN values in a least squares sense \cite{kximp}.
In our case, the end goal with this processing is to have a single $(x, y)$ pixel location representing each unique blob identified in the labeling scheme.

\subsection{Identifying Dimmer Particles}
The previous steps can successfully identify the majority of particles detected in an image, but on occasion it may be desired to extract detections of particles that are very close to the noise level in the image.  
For these types of detections, the previous steps may be inadequate to autonomously extract them from an image, and an alternative approach is needed.
In particular, the median-filter-flattened image does not remove enough stray light from the image, and using a single noise level does not accurately capture variations in the noise level throughout the image due to variations in stray light.
When attempting to detect these particles, we therefore make the following variations to the above steps.

\begin{figure}[h!]
    \centering
    \includegraphics{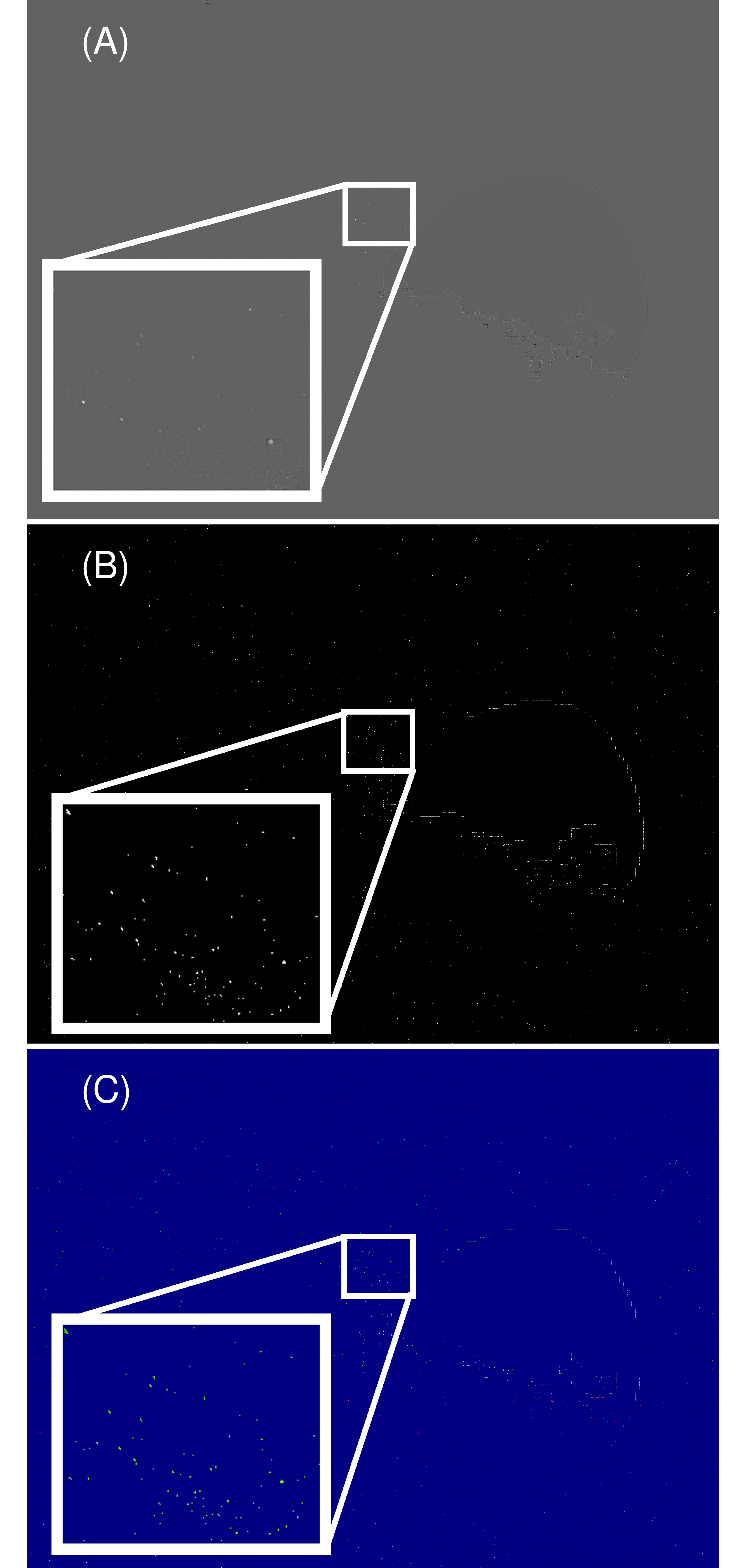}
    \caption{The alternative image processing methods applied to the example image from Fig.~\ref{fig:example_unprocessed}.  (A) The flattened image generated by using the background estimation technique instead of the median filter. (B) The thresholded image generated by using local noise estimates applied to (A). (C) The labeled image generated by applying a connected components algorithm to the binary image in (B).  Isolated blobs of bright pixels have been assigned the same label, and each unique blob shows as a different color.  There are less noisy responses but more possible detections than seen in Fig.~\ref{fig:example_image_processing1} (C). Each sub-figure includes a magnified call-out of the region of interest in the lower left.}
    \label{fig:example_image_processing2}
\end{figure}

First, instead of using a median-filter-flattened image to remove stray light and locate bright spots in the image, we split the image into small subwindows (typically selected to have a size of $15\times15$ pixels so that enough of the background is captured without capturing too many other bright spots in the image) and estimate a background gradient using the DN levels in the subwindow.
Mathematically, this is done by fitting 
\begin{linenomath*}
\begin{equation}
    bg_{ij} = A+Bi+Cj
    \label{eq:background}
\end{equation}
\end{linenomath*}
to the DN values from the subwindows using a least squares estimator where $bg_{ij}$ is the background DN value at pixel $(i,j)$, and $A$, $B$, and $C$ are the coefficients that are fit in the least squares estimator.
Once the background function has been fit, it is evaluated for each pixel in the subwindow and then subtracted from the DN values in the subwindow to produce the flattened DN values for that subwindow.
This approach more accurately represents the variations in the stray light levels that can occur in small windows than does a median filter, thus allowing more of the background signature to be removed and dimmer objects to have a larger signal above the surrounding pixels in the flattened image.
The flattened example image using this technique is shown in Fig.~\ref{fig:example_image_processing2} (A).

Second, instead of using a global noise value for the entire image, the noise level is computed for the same subwindows from which the background was estimated.
This is done by taking the standard deviation of the flattened window.
Using localized noise levels helps to correct for the fact that the noise varies throughout the image.
A global sigma multiplier is still used for the whole image for identifying the detections.
The thresholded image using this alternative technique is shown in Fig.~\ref{fig:example_image_processing2} (B).

Inserting both of these changes into the previous processing steps enables us to detect potential observations that are as low as 2 sigma above the local noise level, without capturing excessive noise.

\subsection{Filtering Particle Detections}
Whether using the general detection techniques or the techniques for extracting dim detections, it is common to have false positive detections that can distract from the real objects.
Therefore, it is desirable to have the ability to filter the presented list of detections to remove some of the false positives.

To begin, we filter detections that correspond to known stars.
This filtering involves matching the detections from the image with a catalogue of known stars \cite<i.e.,>[]{ucac, tycho, gaia}.
There are numerous examples of how to do this in the literature \cite<i.e.,>[]{star1, star2, star3, kximp, astrometry.net}.
Any points that are matched to a known star can safely be ignored from the list of potential particles.
If the star catalogue being used is not complete, there may still be stars contained in the list of potential detections.
To remove these remaining stars, we need to use multiple images and search for potential detections that appear at the same (within the resolution of the detector) inertial right ascension and declination (according to the known camera model and pointing) multiple times.  
When there are observations of the same right ascension and declination in multiple images, there is a high probability that they represent a star that was not successfully matched to a known catalogue star, and these detections can be ignored.
For our analysis we both match points to a star catalogue, as well as use the right ascension and declination matching to ensure as many stars as possible are removed.
In practice, we have found that five detections with the right ascension and declination within 2 pixels of each other over a four-week period indicates a probable star.

The next filtering step is to remove detections that correspond to any extended bodies in the field.  
There are two different techniques that can be used together to filter these detections.
The first technique is to ignore any points whose blobs (found from the connected components labeling) have a large number of points included.
This technique effectively ignores extended bodies in the field of view.
In our analysis, we have found a maximum blob size of 50 square pixels to be a sufficient threshold; however, this is highly dependent on the imager being used and the objects being imaged.

The second technique is to reject detections that are caused by any extended bodies in the field of view, but that are not large enough to be removed by the previous filtering. 
When the terrain of the extended body is mostly smooth, this step is not needed, but when the terrain is rough, as is the case for Bennu, the terminator can cause many false positive detections due to local terrain being illuminated in isolated pockets from the rest of the body (an example of how roughness can lead to false positive detections can be seen in Fig.~\ref{fig:example_unprocessed}).
To reject these points, we need to rely on the knowledge of the relative state between the camera and the extended body, as well as the shape model of the extended body.
Using this knowledge, we can perform a single bounce ray trace, from the camera, through the detection, to the target body's surface, and then to the light source (typically the Sun).
If the ray successfully intersects the body, then does not intersect the body again on the way back to the Sun, we can label the detection as part of the illuminated extended body and thus ignore it.
Due to the uncertainty in the relative state between the camera and the extended body, as well as uncertainty in the shape model of the extended body, this approach will not correctly reject 100 percent of points that are on the extended body, but it typically rejects enough so that further analysis is not overwhelmed with false positives along the terminator.

We now need to remove hot pixels from the list of possible detections.
This can be done by comparing the list of detections with a list of known hot pixels (if one is maintained for the detector) and ignoring detections that are close to a known hot pixel.
If a list of known hot pixels is not maintained for the detector, then we can attempt to identify them using multiple images, similarly to how we identify stars that are not successfully matched to a catalogue.
In this case, we search for pixels that result in a possible detection in multiple images.
When this occurs, it is probably a hot pixel, and the observations can be ignored.
For the NavCam imager used on OSIRIS-REx \cite{tagcams}, a list of known hot pixels is not maintained, therefore we develop the list using the technique described here.
We have found that identifying pixels that have detections in more than 10 images over a four-week period is a useful method of finding hot pixels that can be ignored.

With stars, extended bodies, and hot pixels removed from the list of possible detections, we can now filter based on the properties of the detections themselves.
We use a ``quality code'' that is tuned to work with the specific camera and observation conditions of the OSIRIS-REx proximity operations phases. Different cameras and observing conditions will likely require a modified definition and tuning.

The first component of the quality code is the area of the detection, $A$, defined as the number of connected pixels that are above the specified threshold used in the identification of interesting pixels (the number of the pixels in each blob identified by the connected-components algorithm).
Typically, the area for detections ranges from one square pixel to upwards of 20--30 square pixels, depending on the brightness of the observed object and the threshold used.
In practice, we have found that areas greater than five square pixels are highly likely to be an actual object, with the probability decreasing as the area decreases below this value.

The second component of the quality code is the deviation of the fit PSF from the expected PSF of the camera.
Detections where the fit PSF significantly deviates from the expected PSF are more likely to be false positives than those where the former conforms to the latter.
We measure the deviation of the PSF using the percent difference formula
\begin{linenomath*}
\begin{equation}
    d_{psf}=\frac{\left|\sigma_e-\sigma_m\right|}{\sigma_e}
    \label{eq:psf_deviation}
\end{equation}
\end{linenomath*}
where $d_{psf}$ is the relative deviation from the expected PSF, $\sigma_e$ is the expected semimajor axis of the PSF, and $\sigma_m$ is the measured semimajor axis of the PSF.
We have found that detections with PSF deviations greater than 100 percent are likely false positives.
For NavCam 1, the expected semimajor axis of the PSF is 0.65 pixels.

The final component of the quality code is the signal-to-noise ratio of the detection, computed using the formula in \citeA{ccd_astronomy}
\begin{linenomath*}
\begin{equation}
    s=\frac{\sum_i {DN_f}_i}{\sqrt{\sum_i({DN_f}_i+bg_i) + n(dc+rn)}}
    \label{eq:snr}
\end{equation}
\end{linenomath*}
where $s$ is the signal-to-noise ratio, $\sum_i {DN_f}_i$ is the summed signal of the detection with background removed in electrons, $\sum_i({DN_f}_i+{bg}_i)$ is the summed signal of the detection and the background in electrons, $n$ is the number of pixels that are summed, $dc$ is the dark current per pixel according to the fit dark current profile for the detector, and $rn$ is the read noise of the detector per pixel.
We have found that summing over a $5\times5$ grid of pixels surrounding the peak of the detection provides a reliable measure of the overall signal-to-noise ratio of the detection.
In addition, we have found that detections with signal-to-noise ratios higher than 15 are likely to be real, with the probability of a false detection increasing as the signal-to-noise ratio decreases.

\begin{figure}[h!]
    \centering
    \includegraphics{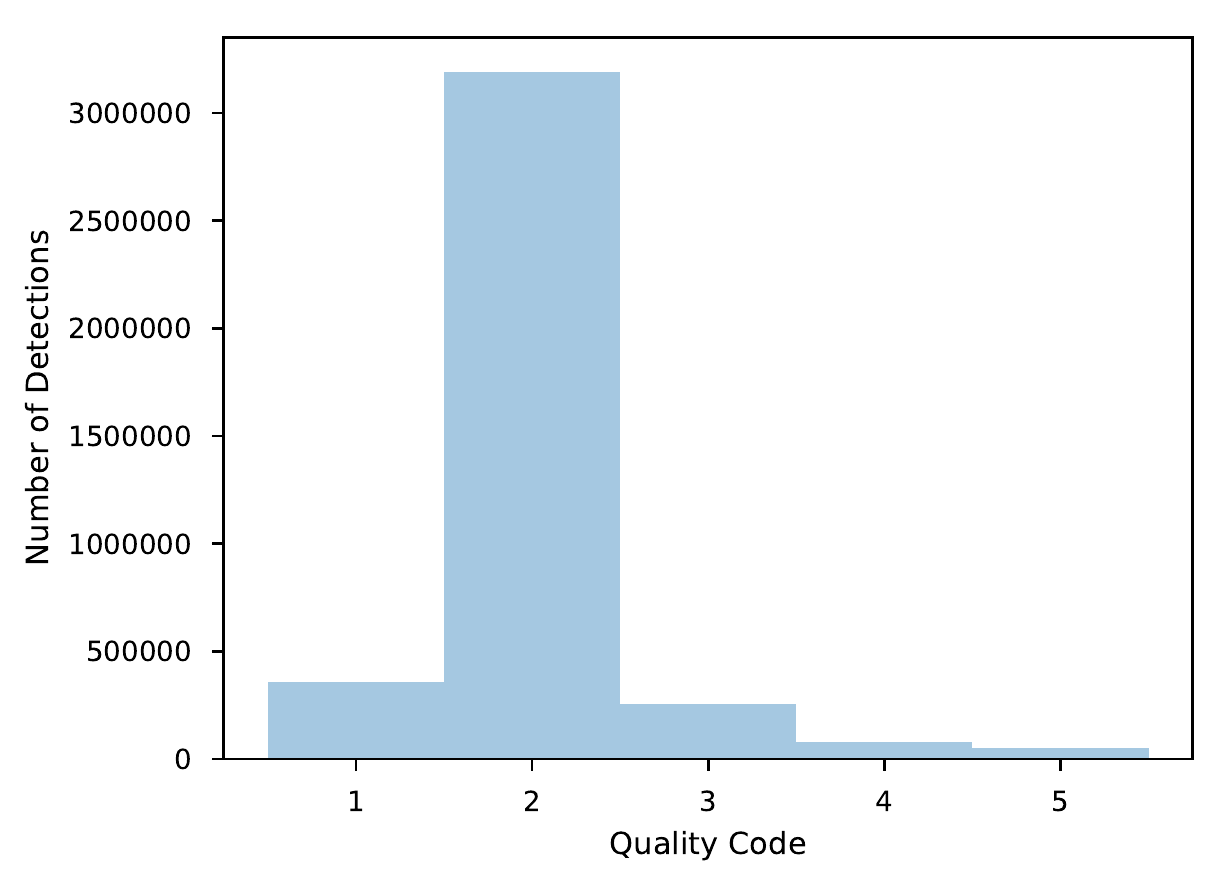}
    \caption{A histogram of the quality code values for detections extracted from images captured by OSIRIS-REx's NavCam 1 camera between 1 January 2019 and 20 July 2019.}
    \label{fig:qual_hist}
\end{figure}

Taking these three components, we can compute the quality of a detection using 
\begin{linenomath*}
\begin{equation}
    q=\frac{\text{clip}(A, 1, 5) + \left(5-4/3\text{clip}(d_{psf}, 0, 3)\right) + \text{clip}(s/3, 1, 5)}{3}
    \label{eq:q_code}
\end{equation}
\end{linenomath*}
where $q$ is the quality, $A$ is the area of the detection, $\text{clip}(x, y, z)$ clips value $x$ to be between $y$ and $z$ such that
\begin{linenomath*}
\begin{equation}
    \text{clip}(x, y, z) = \left\{\begin{array}{cl} y & \text{if } x <y\\x & \text{if } y \leq x \leq z\\z &\text{if } x > z\end{array}\right.
\end{equation}
\end{linenomath*}
and all else is as defined previously.
A quality of 5 indicates that the detection is highly likely to be real, while a quality of 1 indicates that the detection is likely a false positive.
In practice, quality values greater than 4 provide a set of data that is adequate for most human and automated inspections, while detailed analysis of dimmer detections may require quality values between 2 and 3.
An example distribution of the quality codes for detections between 1 January 2019 and 3 May 2019 from NavCam 1 images are shown in Fig.~\ref{fig:qual_hist}.

\section{Tracking of Particles Across Image Frames}
With possible detections extracted for each image, the next step is to identify detections of the same object in multiple frames.
This correspondence confirms whether detections are real objects or false positives and allows for more detailed analysis including event reconstruction \cite{pelgrift}, trajectory fitting \cite{leonard}, and other scientific investigations \cite{lauretta}.
We consider two primary ways that this information can be used to identify objects traversing multiple frames: (1) by generating visual aids for manual linking and (2) by providing detection data to automated linking algorithms.
We discuss both cases here.

\subsection{Visual Aids for Manual Linking of Particles}
One of the most powerful ways to identify the same object in multiple frames is through manual inspection.
Typically, this process involves ``blinking'' images from frame to frame so that a human analyst can visually track moving objects in each frame.
This becomes more difficult when the orientation of the camera changes dramatically from frame to frame, the movement of a particle from frame to frame is large over the time scale of the observations, or there are numerous objects in each frame that need to be tracked.
Using the possible detections extracted from each image, it becomes possible to create visual aids to make this process easier and faster.

\begin{figure}[h!]
    \centering
    \includegraphics{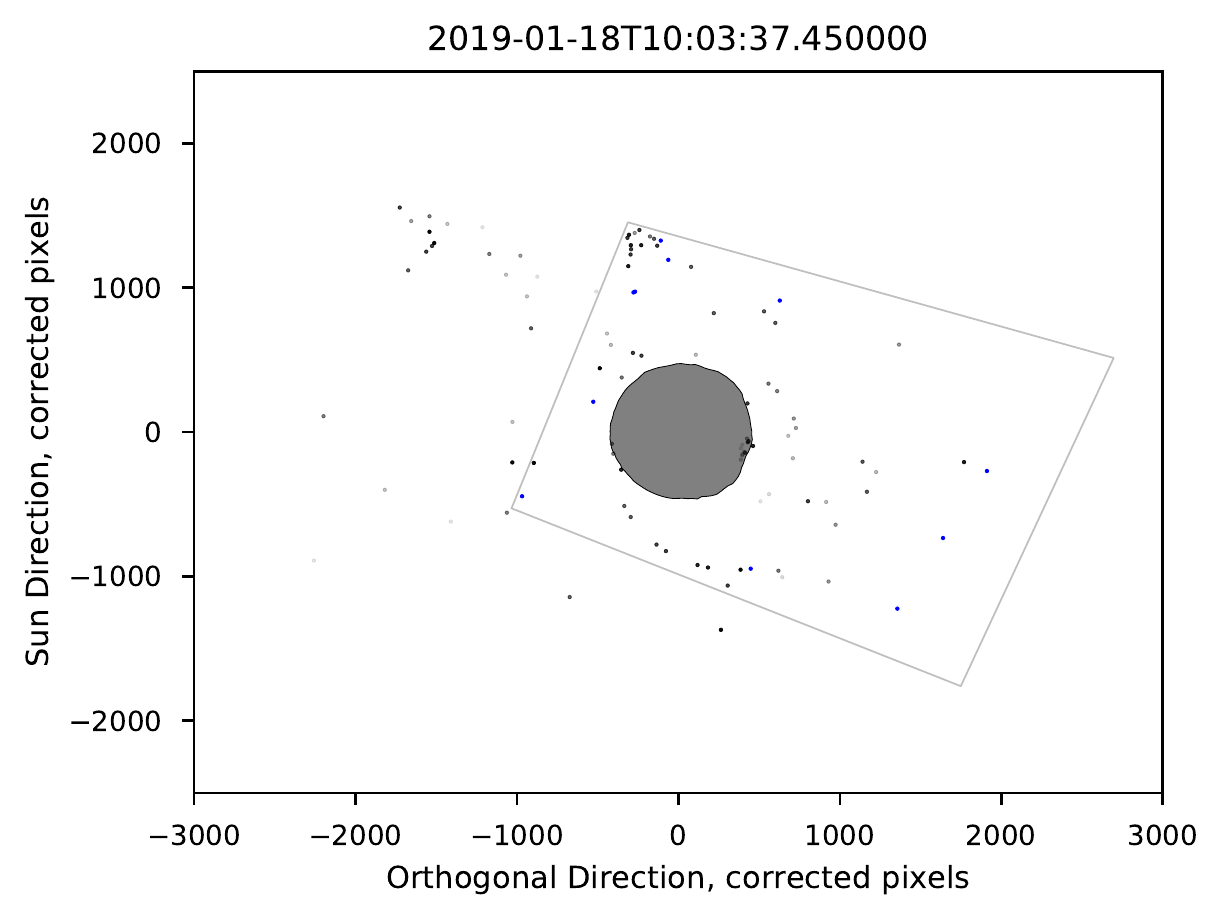}
    \caption{An example of the visual aids created using the autonomously extracted detections for tracking objects manually from frame to frame.  Points in blue correspond to detections from the current frame; points in gray represent detections in previous frames, with detections occurring further from the current frame being more transparent; the gray box represents the field of view of the camera in the current frame; and the outline of the central body is shown as a point of reference.  The points are shown in units of corrected pixels (assuming a pinhole projection camera with no lens distortion modeled after NavCam1) assuming the camera was pointed nadir toward the center-of-mass of Bennu.  There is one evident track immediately to the right of Bennu and one evident track immediately below Bennu.}
    \label{fig:example_inspection_frame}
\end{figure}

To create a visual aid, we begin by taking all potential detections from images occurring less than four hours before the current image frame under consideration.
For each detection, we (1) assume that the object resulting in the detection occurs at the same distance as the central body from the camera along the line-of-sight through the detection, (2) project this point onto a nadir-pointed image plane at the current time using the known camera orientation and the known camera model \cite{camera_model}, and (3) recenter the current image frame at the projected location of the central body.
We then fade each detection according to how long it occurred before the current frame under consideration.
For objects moving from frame to frame, this creates an easily distinguishable trail that is particularly apparent when blinking the visual aids from frame to frame.
An example of one of these frames is shown in Fig.~\ref{fig:example_inspection_frame}.

\subsection{Automated Tracking of Particles}
Even with the visual aids just described, it can be infeasible to manually track particles when there are thousands of frames that need to be considered.
For this reason, we implemented a method for autonomously linking particles from image to image using extended Kalman filters (EKF) \cite{kalman, multi_ekf}.
This technique has proven capable of identifying correspondences between images with only a small number of incorrectly linked detections.
As an added benefit, this technique also provides a guess at the initial state for the particles that can then be fed to other software for more detailed trajectory analysis.
We next provide an overview of the algorithm and the EKF that was implemented, and we show example results of the linking process.

\subsubsection{Automated Tracking Algorithm Overview}
\begin{figure}[h!]
    \centering
    \includegraphics[width=0.75\textwidth]{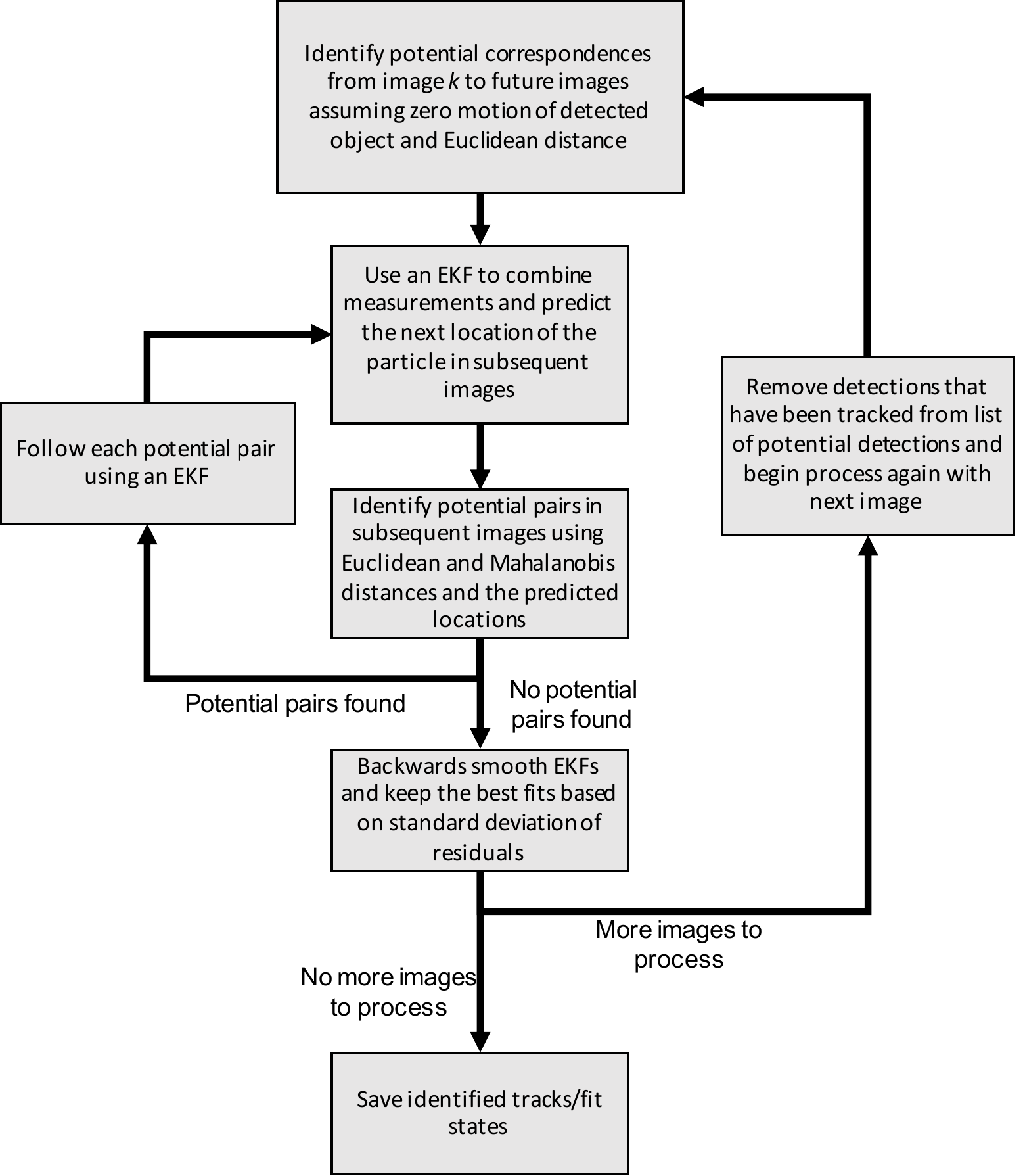}
    \caption{The process of the automated tracking algorithm.  At the end of the algorithm, tracks have been identified and initial orbit determination has been performed on the tracks using the EKFs.}
    \label{fig:flowchart}
\end{figure}

For the automated tracking of particles, we use a multiple object tracking algorithm to simultaneously link and fit trajectories to particles from frame to frame \cite{multi_ekf}.  
For each image $k$, we try to identify any tracks that begin in that image, while ignoring tracks that have already been identified as starting in previous images.
This is done by using an EKF to predict the location of an object in future images, based on the information about that object from observations identified between the starting image $k$ and the current image $k + l$.
These steps are repeated for each image.
A broad overview of the algorithm is shown in the flow chart in Fig.~\ref{fig:flowchart}. More detailed descriptions of select steps are discussed in the following subsections.

We stress that we are forming one EKF for each potential path forward through the images.
As an example, consider a particle observed in the first image (image 1) that has two potential initial pairs in the next image (image 2).  
After creating an EKF to follow each track (two EKFs), the EKFs are propagated to the next image (image 3), and the pairing process described later is performed to identify the next candidate points.
This pairing process finds three potential points in image 3 for the first EKF and two potential points in image 3 for the second EKF.  
The EKFs are cloned to follow each path, resulting in five total EKFs to propagate to the next image (image 4).
This process is repeated until there are no future images to consider, or there are no potential pairs in the remaining images that meet the requirements of the pairing algorithm.
After all the EKFs have reached a terminal point, they are backwards-smoothed to identify which followed viable paths forward according to the dynamics (filtered based on the post-fit standard deviation of the measurement residuals).
These EKFs are kept, and the observations that they include are removed from the pool of potential observations.
The process begins again, but starting with image 2 instead of image 1.

\subsubsection{Automated Tracking Initial Correspondence} \label{sec:initial_corr}
At the beginning of processing for each new image, we do not know what the appropriate correspondences between the detections in the current image and the detections in the future images are, and we do not know how the particles corresponding to the detections in the current image are moving.
Therefore, to do the initial correspondences, we need to allow for a wide range of potential pairings.
To do this, we assume that the particle that resulted in each detection in the current image is located along the observed line-of-sight vector at the same distance from the camera as the central body.
We then assume that the particle is stationary at this point, and we use the known camera attitudes, camera positions, and camera model to determine where this point in inertial space would appear in subsequent images.
In each of the subsequent images, we compute the Euclidean distance between the projected particle locations from the first image and the observed particle locations from the image under consideration, and we accept all potential pairs that meet a specified threshold in pixels.
Because we know that the particles are not stationary, and therefore the projected locations are not actually where the particles will appear in the subsequent images, we need to set the distance threshold to be large, but we do not want it to be so large that there are extreme numbers of potential pairings to track.
In our case, we found a threshold of 400 pixels to be effective for this initial pairing, except in exceedingly crowded particle fields (such as an ejecta event) where a tighter initial distance of 200 pixels was used.
We perform this initial pairing between points in all images captured less than 1 hour later than the first image.
We are only looking for tracks that begin in the image under consideration.  
Therefore, any detections that have already been linked to tracks that start in previous images are ignored at this stage.

\subsubsection{Extended Kalman Filter Design and Tuning}
With the initial pairings made between the first image and subsequent images, we need a way to follow these potential tracks and determine whether we can link more observations to them.
An EKF is ideally suited to this task because the prediction of the particle locations in future images improves as more valid observations are added to the track. It is also computationally efficient.
EKFs are well described in the literature \cite{kalman, multi_ekf}. 
Here we describe specific implementation details that we used in our processing.

\paragraph{Extended Kalman Filter Dynamics}
At the heart of the EKF is the dynamics model used to predict the current state to a subsequent time.
That dynamical model needs to be accurate enough to allow useful predictions, but simple enough that the EKF can run in a reasonable time, especially given that we are processing many different EKFs to identify the tracks.
We are tracking objects that are orbiting a small body, with a small gravity field.  
Therefore, there are two dominant forces acting on the particles: the gravity field of the central body and the solar radiation pressure.
We include the gravity as a point mass model and the solar radiation pressure as a basic cannonball model.
The gravity assumes the estimated values obtained from the spacecraft's operations in proximity to the body \cite{bennu_gravity}. 
A scaling term is estimated as part of the state vector for the solar radiation pressure because the particles may have different sizes and masses.

\paragraph{Extended Kalman Filter State Vector}
The state vector represents all of the information needed to describe the object being tracked and to predict to future time steps using the dynamics model.
For our processing, the state vector includes the inertial position and velocity of the particle with respect to Bennu, as well as a solar radiation term to estimate the effects of solar radiation on the particles, leading to
\begin{linenomath*}
\begin{equation}
    \mathbf{x}=\left[\begin{array}{ccccccc} p_x & p_y & p_z & v_x & v_y & v_z & s\end{array}\right]
    \label{eq:state}
\end{equation}
\end{linenomath*}
where $p_{x-z}$ are the components of the inertial position, $v_{x-z}$ are the components of the inertial velocity, and $s$ is the solar radiation term assuming a cannonball model \cite{solrad}.
The solar radiation term represents the combination of the surface area of the object, the coefficient of reflection, and the mass of the object 
\begin{linenomath*}
\begin{equation}
    s=\frac{c_rA}{m}
    \label{eq:cram}
\end{equation}
\end{linenomath*}
where $c_r$ is the coefficient of reflection, $A$ is the surface area of the particle, and $m$ is the mass of the particle.
To initialize the state vector, we assume that the object is located along the line-of-sight vector through the observation at the same distance as the central body is from the camera; we further assume that the initial velocity with respect to Bennu is zero and the solar radiation term is 0.08 m\textsuperscript{2}/kg.

\paragraph{Extended Kalman Filter Measurements}
The measurements used to update the EKF are the observations of the particles in each image, expressed in pixels.  
The known attitude of the camera and the known geometric camera model are used to transform the state to the predicted observation at a given time.
Partial derivatives of the camera model and rotation were derived to compute the measurement sensitivity matrix, used to linearly predict how a change in the state would change the predicted particle location and the resulting residual.  

\paragraph{Extended Kalman Filter Tuning}
EKFs need to be tuned to be effective.
Aspects that should be tuned include the uncertainty on the initial state, the weighting assumed for each measurement, and the process noise included to account for deviations from the assumed dynamics.
We assumed the measurements were each accurate to one tenth of a pixel ($1\sigma$), the uncertainty in the initial position was one tenth of a pixel in plane of sky and 5 km in the range direction ($1\sigma$), the initial uncertainty in the velocity was 50 cm/s ($1\sigma$), and the solar radiation term initial uncertainty was 1 m\textsuperscript{2}/kg ($1\sigma$).
For the process noise, we used a weighting scheme based on the range to the body, because the closer to the body the particle is, the greater the effects of the gravity mis-modeling from assuming a point mass.
In addition, we kept the process noise large to reflect the overall uncertainty in the dynamical environment.
We used a process noise of
\begin{linenomath*}
\begin{equation}
    \mathbf{Q}=\left[\begin{array}{ccccccc} 
            0 & 0 & 0 & 0 & 0 & 0 & 0 \\
            0 & 0 & 0 & 0 & 0 & 0 & 0 \\
            0 & 0 & 0 & 0 & 0 & 0 & 0 \\
            0 & 0 & 0 & q^2 & 0 & 0 & 0 \\
            0 & 0 & 0 & 0 & q^2 & 0 & 0 \\
            0 & 0 & 0 & 0 & 0 & q^2 & 0 \\
            0 & 0 & 0 & 0 & 0 & 0 & 0
    \end{array} \right]
\end{equation}
\end{linenomath*}
where $q=-1.816\times10^{-6}(r-250)+1\times10^{-3}$ m/s/s\textsuperscript{1/2}, $r$ is the radial distance from the central body in meters, and $q$ has a minimum value of $1\times10^{-6}$ m/s/s\textsuperscript{1/2} at a range of 800 m.
The process noise matrix is added directly to the time derivative of the covariance matrix using
\begin{linenomath*}
\begin{equation}
    \dot{\mathbf{P}}=\mathbf{F}\mathbf{P} + \mathbf{P}\mathbf{F}^T + \mathbf{Q}
    \label{eq:pdot}
\end{equation}
\end{linenomath*}
where $\dot{\mathbf{P}}$ is the covariance time derivative and $\mathbf{F}$ is the Jacobian matrix of the dynamics model with respect to a change in the state vector.
This increases the velocity uncertainty between 0.7 mm/s and 70 cm/s ($1\sigma$) over the course of 30 minutes depending on the range to the body.
This high level of process noise (particularly near the surface) helps to keep the filters from becoming locked onto a poor solution and has proven useful for fitting a range of particle trajectories.

\subsubsection{Subsequent Detection Pairings}
After the first pairs of detections are created (Section \ref{sec:initial_corr}), EKFs are started for each pair.  
From this point on, the EKF is used to predict the location of the next observation in upcoming images.  
The predicted location from the EKF is then compared with the observed detections for the image under consideration using the Euclidean and Mahalanobis distance \cite{mahalanobis}.
If both the distance thresholds are met between the predicted location and the observed locations, they are considered a potential pair.
For each potential pair, a new EKF is cloned from the detecting EKF to follow that track.
The cloning process includes copying the current state and covariance up to the point of the cloning, as well as the history of all previous measurements ingested by the EKF to this point.
For each clone, the paired measurement is provided as the next measurement, and the process continues until no more potential pairs meeting the threshold are found.

When performing this pairing, images with time stamps within one hour of the last image used to update the EKF are all considered.  
Therefore, a single EKF may have potential pairs from more than one image to clone and follow for the next step.  
This thoroughness helps to ensure that the correct path is followed.

The Euclidean distance threshold used in this step to identify possible pairings is determined by anticipating that the more measurements that have been ingested by the EKF, the better the prediction will be.
We implemented the following linear distance threshold
\begin{linenomath*}
\begin{equation}
    d_{thresh}=50(8-n)+10
    \label{eq:distance_threshold}
\end{equation}
\end{linenomath*}
where $d_{thresh}$ is the distance threshold in pixels, $n$ is the number of measurements that have been ingested by the EKF, and $d_{thresh}$ has a minimum value of 10 pixels.
In addition to the Euclidean threshold, we use a Mahalanobis threshold, given by 
\begin{linenomath*}
\begin{equation}
    m^2=\left(\mathbf{y}-\hat{\mathbf{y}}\right)^T\left(\mathbf{P}_{\mathbf{x}}^I+\mathbf{R}\right)^{-1}\left(\mathbf{y}-\hat{\mathbf{y}}\right)
    \label{eq:mahal}
\end{equation}
\end{linenomath*}
with $m^2$ being the squared Mahalanobis distance, $\mathbf{y}$ being the pixel location of the current observation under consideration, $\hat{\mathbf{y}}$ being the predicted location of the observation in the current image, $\mathbf{P}_{\mathbf{x}}^I$ being the covariance of the state vector mapped into the measurement space, and $\mathbf{R}$ being the covariance matrix of the observation itself in the measurement space \cite{mahalanobis}.
In our analysis, we have excluded potential paths where the next detection has a Mahalanobis squared distance of more than 25.
Using both the Euclidean distance and the Mahalanobis distance makes for a more robust system.
The Euclidean distance threshold helps to ensure that the right path is followed early on, before the filter has converged on a solution, while the Mahalanobis distance helps to reject false paths once the filter has converged.

This process is repeated until every EKF started from the first image has either run out of future images to consider, or has reached a point where no more potential pairs meet the distance thresholds.

\subsubsection{Backwards Smoothing, Identification of Valid Tracks, and Next Images}
Once the linking process is complete and the EKFs have been linked to the end of the available images, or to a point where the distance thresholds are no longer met by any future points, the EKFs are backwards smoothed to generate the best fit trajectory through all of the measurements.  
The residuals from this backward smoothing are then used to determine which EKFs found valid tracks.
We find that valid tracks have smoothed residuals with a standard deviation less than 25 pixels ($1\sigma$); therefore, any tracks whose residuals meet this threshold are considered to be real tracks.
The detections from these tracks are removed from the list of possible detections, and the next image is fed through the same process to identify tracks that start in it.
The process is repeated to identify tracks that begin in each image until all images are processed.

\subsubsection{Results from Automated Tracking of Particles}
\begin{figure}[htbp]
    \centering
    \includegraphics{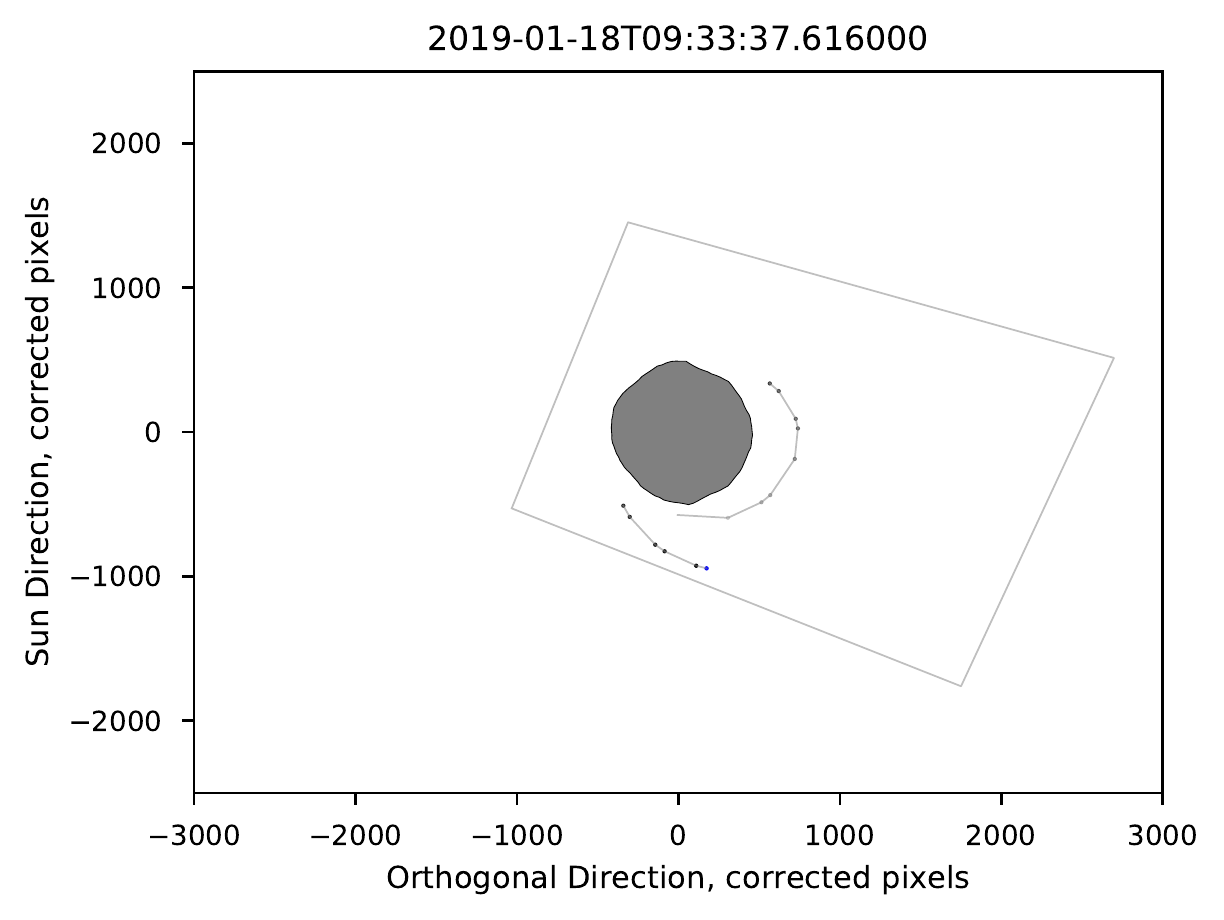}
    \caption{An example of the tracks found by the automated algorithm from images taken on 18 January 2019.  The tracks are shown in gray, with the observations belonging to the track shown as a scatter (blue and gray).}  
    \label{fig:automated_tracks1}
\end{figure}

\begin{figure}[htbp]
    \centering
    \includegraphics{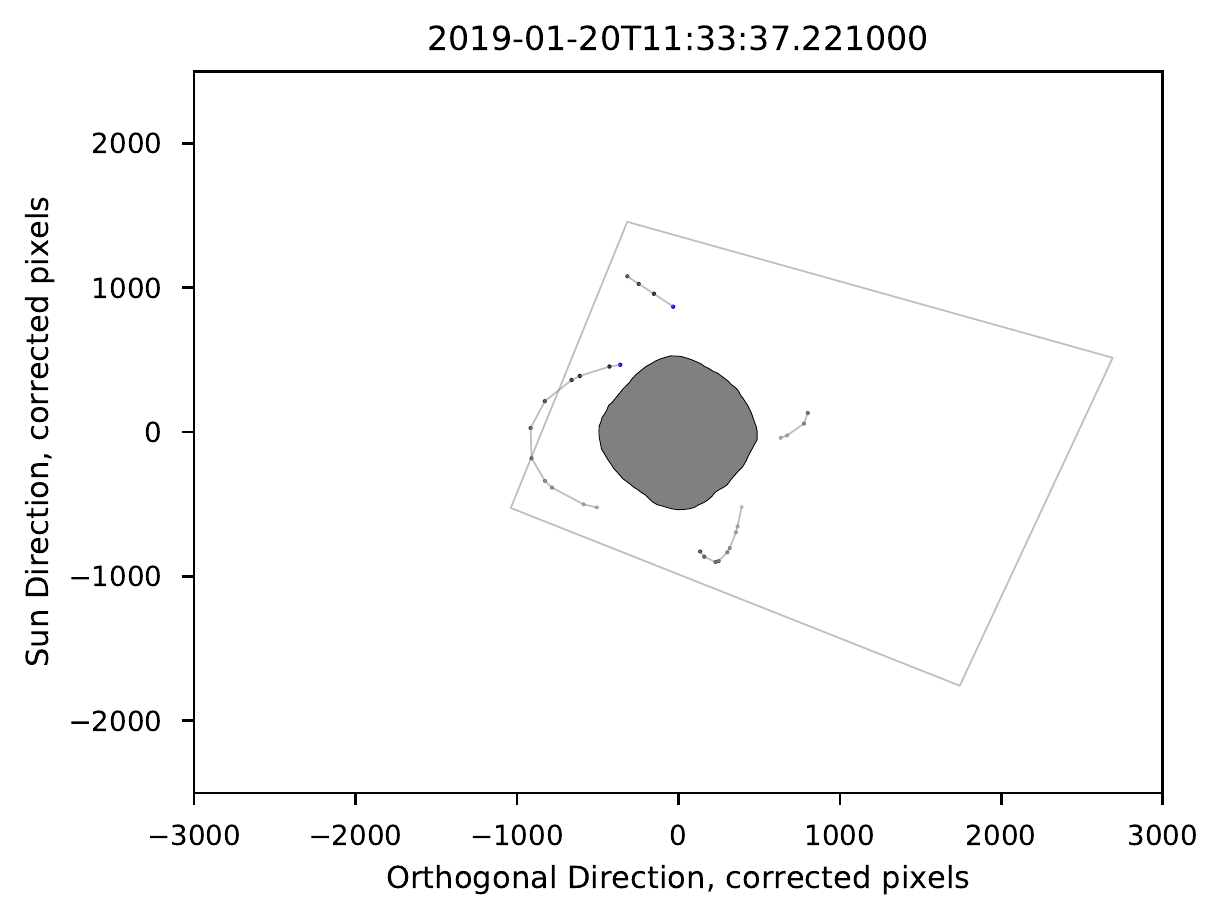}
    \caption{An example of the tracks found by the automated algorithm from images taken on 20 January 2019.  The tracks are shown in gray with the observations belonging to the track shown as a scatter (blue and gray).} 
    \label{fig:automated_tracks2}
\end{figure}

\begin{figure}[htbp]
    \centering
    \includegraphics{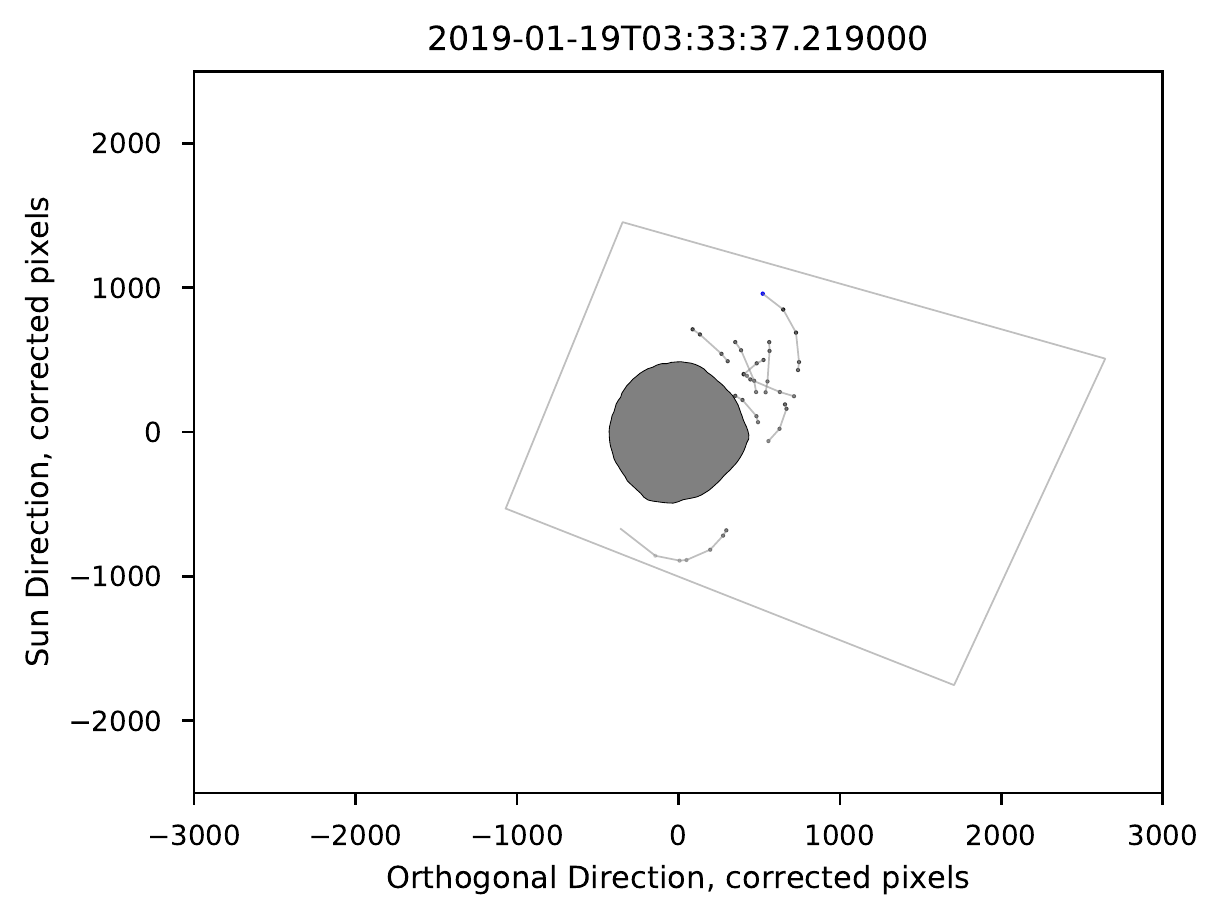}
    \caption{An example of the performance of the automated algorithm when the number of possible detections is crowded, as during an ejection event, from images taken on 19 January 2019.  In the area of the ejection (to the top right of Bennu), a couple of false positive tracks are identified, but there are also some correct identifications.}
    \label{fig:crowded1}
\end{figure}

\begin{figure}[htbp]
    \centering
    \includegraphics{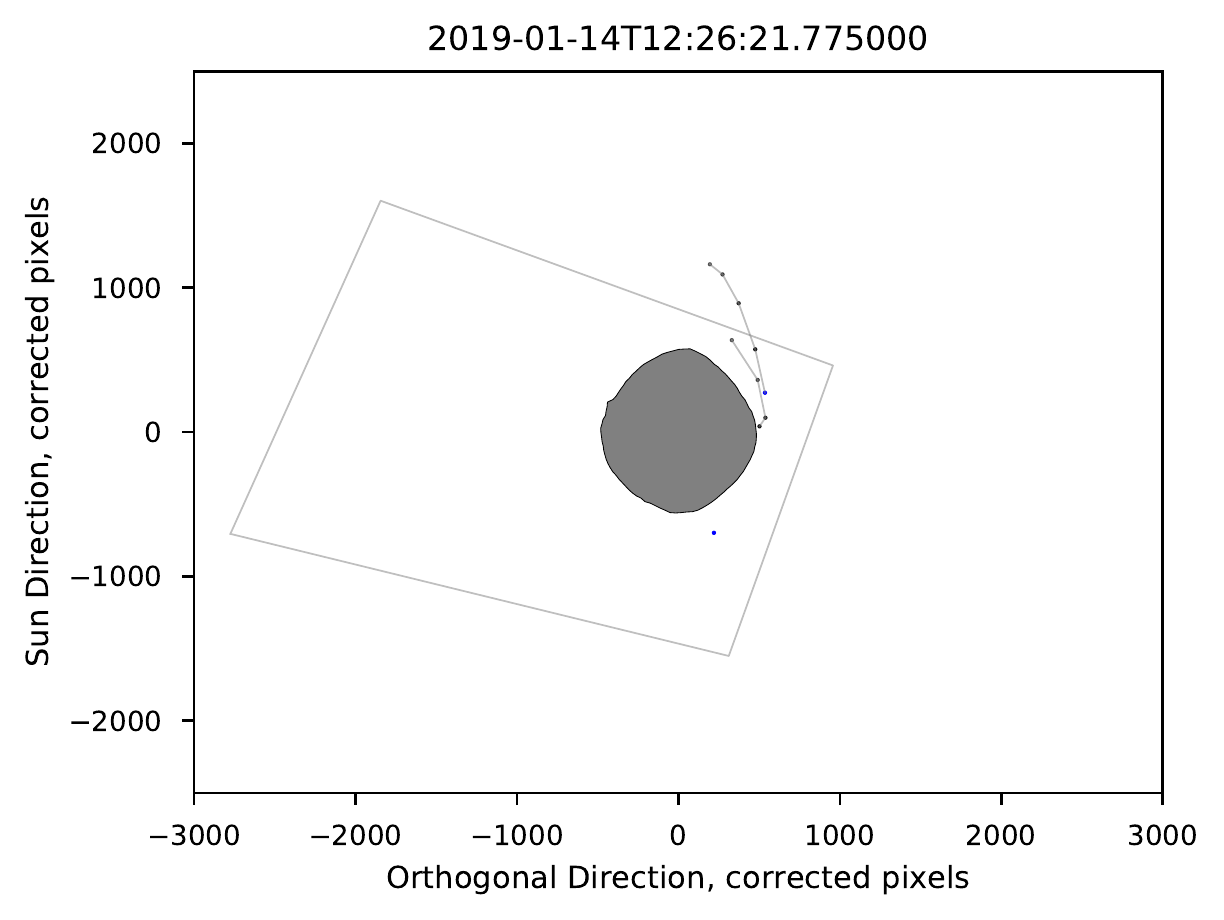}
    \caption{An example of how a false positive can interfere with what is actually a real track, from images taken on 14 January  2019 (to the top right of Bennu).  Because detections that have already been placed in a track are ignored, real tracks cannot use these detections again, even if they are part of the track.  Rejecting the false positives will allow for fewer false negatives.}  
    \label{fig:false_positive}
\end{figure}

Applying the above algorithms to images captured by NavCam 1 while the OSIRIS-REx spacecraft was in orbit during January 2019 successfully produced 103 tracks.
Manual tracking was also performed on these images, resulting in 95 tracks.
The automated process successfully found 70 percent of the manually identified tracks, and also found 38 additional tracks that were not originally detected manually but were identified after observing the output of the automatic track generation tool.  
Some false positive tracks were detected, particularly near the location where the terminator and illuminated limb meet, due to a high number of false positive detections from the image processing in this region.
We are currently investigating techniques to reject these false positives.
The majority of the tracks that were not identified were caused by false positive identifications interfering with the track's path; therefore, finding a way to filter false positives should allow an even higher percentage of all tracks to be identified.
Examples of the tracks identified by the automated algorithm are shown in Figs.~\ref{fig:automated_tracks1} and \ref{fig:automated_tracks2}.
Figs.~\ref{fig:crowded1} and \ref{fig:false_positive} show what happens with some of the false positive identifications.


\section{Conclusion}
We describe autonomous methods to extract and link detections of particles in orbit around a celestial body, as observed by a monocular camera.
We demonstrate these techniques using images of particles ejected from the surface of the asteroid Bennu, acquired in January 2019 by the NavCam 1 camera onboard the OSIRIS-REx spacecraft. 
Our methods decrease the amount of manual effort that is required to examine these images and directly enable analysis of the ejected particles and their implications for Bennu.
These algorithms are being applied to all images captured by OSIRIS-REx during proximity operations about Bennu so that longer-term trends in the frequency of the particles can be examined.

\acknowledgments
The authors are grateful to the entire OSIRIS-REx Team for making the encounter with Bennu possible.
They would also like to thank Natalie Liounis and Cat Wolner for their assistance in preparing this document.
This material is based upon work supported by NASA under Contracts NNM10AA11C and NNG13FC02C issued through the New Frontiers Program. 
A portion of this work was conducted at the Jet Propulsion Laboratory, California Institute of Technology under a contract with the National Aeronautics and Space Administration.
Image data will be available via the Planetary Data System (PDS) (https://{\allowbreak}sbn.psi.edu/{\allowbreak}pds/{\allowbreak}resource/{\allowbreak}orex/). 
Data are delivered to the PDS according to the OSIRIS-REx Data Management Plan available in the OSIRIS-REx PDS archive. 
Higher-level products including possible detections and tracks will be available in the Planetary Data System 1 year after departure from the asteroid in accordance with the OSIRIS-REx Data Management Plan.


\bibliography{particles}

\begin{thebibliography}{}

\bibitem [\protect \citeauthoryear {%
{Al-Shakarji}%
, {Seetharaman}%
, {Bunyak}%
\BCBL {}\ \BBA {} {Palaniappan}%
}{%
{Al-Shakarji}%
\ \protect \BOthers {.}}{%
{\protect \APACyear {2017}}%
}]{%
multi_ekf}
\APACinsertmetastar {%
multi_ekf}%
\begin{APACrefauthors}%
{Al-Shakarji}, N\BPBI M.%
, {Seetharaman}, G.%
, {Bunyak}, F.%
\BCBL {}\ \BBA {} {Palaniappan}, K.%
\end{APACrefauthors}%
\unskip\
\newblock
\APACrefYearMonthDay{2017}{}{}.
\newblock
{\BBOQ}\APACrefatitle {{Robust Multi-Object Tracking with Semantic Color
  Correlation}} {{Robust Multi-Object Tracking with Semantic Color
  Correlation}}.{\BBCQ}
\newblock
\BIn{} \APACrefbtitle {{2017 14th IEEE International Conference on Advanced
  Video and Signal Based Surveillance (AVSS)}} {{2017 14th IEEE International
  Conference on Advanced Video and Signal Based Surveillance (AVSS)}}\
  (\BPG~1-7).
\PrintBackRefs{\CurrentBib}

\bibitem [\protect \citeauthoryear {%
Bos%
\ \protect \BOthers {.}}{%
Bos%
\ \protect \BOthers {.}}{%
{\protect \APACyear {2018}}%
}]{%
tagcams}
\APACinsertmetastar {%
tagcams}%
\begin{APACrefauthors}%
Bos, B.%
, Ravine, M.%
, Caplinger, M.%
, Schaffner, J.%
, Ladewig, J.%
, Olds, R.%
\BDBL {}others%
\end{APACrefauthors}%
\unskip\
\newblock
\APACrefYearMonthDay{2018}{}{}.
\newblock
{\BBOQ}\APACrefatitle {{Touch and Go Camera System (TAGCAMS) for the OSIRIS-REx
  Asteroid Sample Return Mission}} {{Touch and Go Camera System (TAGCAMS) for
  the OSIRIS-REx Asteroid Sample Return Mission}}.{\BBCQ}
\newblock
\APACjournalVolNumPages{Space Science Reviews}{214}{1}{37}.
\PrintBackRefs{\CurrentBib}

\bibitem [\protect \citeauthoryear {%
Brown%
\ \protect \BOthers {.}}{%
Brown%
\ \protect \BOthers {.}}{%
{\protect \APACyear {2018}}%
}]{%
gaia}
\APACinsertmetastar {%
gaia}%
\begin{APACrefauthors}%
Brown, A.%
, Vallenari, A.%
, Prusti, T.%
, De~Bruijne, J.%
, Babusiaux, C.%
, Bailer-Jones, C.%
\BDBL {}others%
\end{APACrefauthors}%
\unskip\
\newblock
\APACrefYearMonthDay{2018}{}{}.
\newblock
{\BBOQ}\APACrefatitle {{Gaia Data Release 2-Summary of the Contents and Survey
  Properties}} {{Gaia Data Release 2-Summary of the Contents and Survey
  Properties}}.{\BBCQ}
\newblock
\APACjournalVolNumPages{{Astronomy \& Astrophysics}}{616}{}{A1}.
\PrintBackRefs{\CurrentBib}

\bibitem [\protect \citeauthoryear {%
{Di Stefano}%
\ \BBA {} {Bulgarelli}%
}{%
{Di Stefano}%
\ \BBA {} {Bulgarelli}%
}{%
{\protect \APACyear {1999}}%
}]{%
connected}
\APACinsertmetastar {%
connected}%
\begin{APACrefauthors}%
{Di Stefano}, L.%
\BCBT {}\ \BBA {} {Bulgarelli}, A.%
\end{APACrefauthors}%
\unskip\
\newblock
\APACrefYearMonthDay{1999}{}{}.
\newblock
{\BBOQ}\APACrefatitle {{A Simple and Efficient Connected Components Labeling
  Algorithm}} {{A Simple and Efficient Connected Components Labeling
  Algorithm}}.{\BBCQ}
\newblock
\BIn{} \APACrefbtitle {{Proceedings 10th International Conference on Image
  Analysis and Processing}} {{Proceedings 10th International Conference on
  Image Analysis and Processing}}\ (\BPG~322-327).
\PrintBackRefs{\CurrentBib}

\bibitem [\protect \citeauthoryear {%
Gehrels%
}{%
Gehrels%
}{%
{\protect \APACyear {1991}}%
}]{%
modp1}
\APACinsertmetastar {%
modp1}%
\begin{APACrefauthors}%
Gehrels, T.%
\end{APACrefauthors}%
\unskip\
\newblock
\APACrefYearMonthDay{1991}{}{}.
\newblock
{\BBOQ}\APACrefatitle {{Scanning with Charge-Coupled Devices}} {{Scanning with
  Charge-Coupled Devices}}.{\BBCQ}
\newblock
\APACjournalVolNumPages{Space Science Reviews}{58}{1}{347--375}.
\PrintBackRefs{\CurrentBib}

\bibitem [\protect \citeauthoryear {%
Groeneveld%
\ \BBA {} Kuiper%
}{%
Groeneveld%
\ \BBA {} Kuiper%
}{%
{\protect \APACyear {1954}}%
}]{%
blinking}
\APACinsertmetastar {%
blinking}%
\begin{APACrefauthors}%
Groeneveld, I.%
\BCBT {}\ \BBA {} Kuiper, G\BPBI P.%
\end{APACrefauthors}%
\unskip\
\newblock
\APACrefYearMonthDay{1954}{}{}.
\newblock
{\BBOQ}\APACrefatitle {{Photometric Studies of Asteroids. I}} {{Photometric
  Studies of Asteroids. I}}.{\BBCQ}
\newblock
\APACjournalVolNumPages{The Astrophysical Journal}{120}{}{200}.
\PrintBackRefs{\CurrentBib}

\bibitem [\protect \citeauthoryear {%
Helin%
, Pravdo%
, Rabinowitz%
\BCBL {}\ \BBA {} Lawrence%
}{%
Helin%
\ \protect \BOthers {.}}{%
{\protect \APACyear {1997}}%
}]{%
neat}
\APACinsertmetastar {%
neat}%
\begin{APACrefauthors}%
Helin, E\BPBI F.%
, Pravdo, S\BPBI H.%
, Rabinowitz, D\BPBI L.%
\BCBL {}\ \BBA {} Lawrence, K\BPBI J.%
\end{APACrefauthors}%
\unskip\
\newblock
\APACrefYearMonthDay{1997}{}{}.
\newblock
{\BBOQ}\APACrefatitle {{Near-Earth Asteroid Tracking (NEAT) Program}}
  {{Near-Earth Asteroid Tracking (NEAT) Program}}.{\BBCQ}
\newblock
\APACjournalVolNumPages{Annals of the New York Academy of
  Sciences}{822}{1}{6-25}.
\PrintBackRefs{\CurrentBib}

\bibitem [\protect \citeauthoryear {%
Hog%
\ \protect \BOthers {.}}{%
Hog%
\ \protect \BOthers {.}}{%
{\protect \APACyear {{2000}}}%
}]{%
tycho}
\APACinsertmetastar {%
tycho}%
\begin{APACrefauthors}%
Hog, E.%
, Fabricius, C.%
, Makarov, V.%
, Urban, S.%
, Corbin, T.%
, Wycoff, G.%
\BDBL {}Wicenec, A.%
\end{APACrefauthors}%
\unskip\
\newblock
\APACrefYearMonthDay{{2000}}{{MAR}}{}.
\newblock
{\BBOQ}\APACrefatitle {{The Tycho-2 Catalogue of the 2.5 Million Brightest
  Stars}} {{The Tycho-2 Catalogue of the 2.5 Million Brightest Stars}}.{\BBCQ}
\newblock
\APACjournalVolNumPages{{ASTRONOMY \& ASTROPHYSICS}}{{355}}{{2}}{{L27-L30}}.
\PrintBackRefs{\CurrentBib}

\bibitem [\protect \citeauthoryear {%
Howell%
}{%
Howell%
}{%
{\protect \APACyear {2006}}%
}]{%
ccd_astronomy}
\APACinsertmetastar {%
ccd_astronomy}%
\begin{APACrefauthors}%
Howell, S\BPBI B.%
\end{APACrefauthors}%
\unskip\
\newblock
\APACrefYear{2006}.
\newblock
\APACrefbtitle {{Handbook of CCD Astronomy}} {{Handbook of CCD Astronomy}}.
\newblock
\APACaddressPublisher{}{Cambridge University Press}.
\PrintBackRefs{\CurrentBib}

\bibitem [\protect \citeauthoryear {%
Iglewicz%
\ \BBA {} Hoaglin%
}{%
Iglewicz%
\ \BBA {} Hoaglin%
}{%
{\protect \APACyear {1993}}%
}]{%
z-score}
\APACinsertmetastar {%
z-score}%
\begin{APACrefauthors}%
Iglewicz, B.%
\BCBT {}\ \BBA {} Hoaglin, D.%
\end{APACrefauthors}%
\unskip\
\newblock
\APACrefYearMonthDay{1993}{}{}.
\newblock
{\BBOQ}\APACrefatitle {{Volume 16: How to Detect and Handle Outliers}} {{Volume
  16: How to Detect and Handle Outliers}}.{\BBCQ}
\newblock
\APACjournalVolNumPages{{The ASQC Basic References in Quality Control:
  Statistical Techniques}}{16}{}{}.
\PrintBackRefs{\CurrentBib}

\bibitem [\protect \citeauthoryear {%
Jackman%
\ \BBA {} Dumont%
}{%
Jackman%
\ \BBA {} Dumont%
}{%
{\protect \APACyear {2013}}%
}]{%
kximp}
\APACinsertmetastar {%
kximp}%
\begin{APACrefauthors}%
Jackman, C.%
\BCBT {}\ \BBA {} Dumont, P.%
\end{APACrefauthors}%
\unskip\
\newblock
\APACrefYearMonthDay{2013}{}{}.
\newblock
{\BBOQ}\APACrefatitle {{Optical Navigation Capabilities for Deep Space
  Missions}} {{Optical Navigation Capabilities for Deep Space
  Missions}}.{\BBCQ}
\newblock
\BIn{} \APACrefbtitle {{Proceedings of the 23rd AAS/AIAA Space Flight Mechanics
  Conference}.} {{Proceedings of the 23rd AAS/AIAA Space Flight Mechanics
  Conference}.}
\PrintBackRefs{\CurrentBib}

\bibitem [\protect \citeauthoryear {%
Kalman%
}{%
Kalman%
}{%
{\protect \APACyear {1960}}%
}]{%
kalman}
\APACinsertmetastar {%
kalman}%
\begin{APACrefauthors}%
Kalman, R\BPBI E.%
\end{APACrefauthors}%
\unskip\
\newblock
\APACrefYearMonthDay{1960}{}{}.
\newblock
{\BBOQ}\APACrefatitle {{A New Approach to Linear Filtering and Prediction
  Problems}} {{A New Approach to Linear Filtering and Prediction
  Problems}}.{\BBCQ}
\newblock
\APACjournalVolNumPages{{Transactions of the ASME--Journal of Basic
  Engineering}}{82}{Series D}{35--45}.
\PrintBackRefs{\CurrentBib}

\bibitem [\protect \citeauthoryear {%
Lang%
, Hogg%
, Mierle%
, Blanton%
\BCBL {}\ \BBA {} Roweis%
}{%
Lang%
\ \protect \BOthers {.}}{%
{\protect \APACyear {2010}}%
}]{%
astrometry.net}
\APACinsertmetastar {%
astrometry.net}%
\begin{APACrefauthors}%
Lang, D.%
, Hogg, D\BPBI W.%
, Mierle, K.%
, Blanton, M.%
\BCBL {}\ \BBA {} Roweis, S.%
\end{APACrefauthors}%
\unskip\
\newblock
\APACrefYearMonthDay{2010}{}{}.
\newblock
{\BBOQ}\APACrefatitle {{Astrometry.net: Blind Astrometric Calibration of
  Arbitrary Astronomical Images}} {{Astrometry.net: Blind Astrometric
  Calibration of Arbitrary Astronomical Images}}.{\BBCQ}
\newblock
\APACjournalVolNumPages{{The Astronomical Journal}}{139}{5}{1782}.
\PrintBackRefs{\CurrentBib}

\bibitem [\protect \citeauthoryear {%
Larsen%
\ \protect \BOthers {.}}{%
Larsen%
\ \protect \BOthers {.}}{%
{\protect \APACyear {2007}}%
}]{%
spacewatch}
\APACinsertmetastar {%
spacewatch}%
\begin{APACrefauthors}%
Larsen, J\BPBI A.%
, Roe, E\BPBI S.%
, Albert, C\BPBI E.%
, Descour, A\BPBI S.%
, McMillan, R\BPBI S.%
, Gleason, A\BPBI E.%
\BDBL {}others%
\end{APACrefauthors}%
\unskip\
\newblock
\APACrefYearMonthDay{2007}{}{}.
\newblock
{\BBOQ}\APACrefatitle {{The Search for Distant Objects in the Solar System
  Using Spacewatch}} {{The Search for Distant Objects in the Solar System Using
  Spacewatch}}.{\BBCQ}
\newblock
\APACjournalVolNumPages{The Astronomical Journal}{133}{4}{1247}.
\PrintBackRefs{\CurrentBib}

\bibitem [\protect \citeauthoryear {%
Lauretta%
, Hergenrother%
\BCBL {}\ \protect \BOthers {.}}{%
Lauretta%
\ \protect \BOthers {.}}{%
{\protect \APACyear {in review}}%
}]{%
lauretta}
\APACinsertmetastar {%
lauretta}%
\begin{APACrefauthors}%
Lauretta, D\BPBI S.%
, Hergenrother, C\BPBI W.%
\BCBL {}\ \BOthersPeriod {.}\end{APACrefauthors}%
\unskip\
\newblock
\APACrefYearMonthDay{in review}{}{}.
\newblock
{\BBOQ}\APACrefatitle {{OSIRIS-REx Discovery of Particle Ejection from Active
  Asteroid (101955) Bennu}} {{OSIRIS-REx Discovery of Particle Ejection from
  Active Asteroid (101955) Bennu}}.{\BBCQ}
\newblock
\APACjournalVolNumPages{Science}{}{}{}.
\PrintBackRefs{\CurrentBib}

\bibitem [\protect \citeauthoryear {%
Leonard%
\ \protect \BOthers {.}}{%
Leonard%
\ \protect \BOthers {.}}{%
{\protect \APACyear {in preparation}}%
}]{%
leonard}
\APACinsertmetastar {%
leonard}%
\begin{APACrefauthors}%
Leonard, J\BPBI M.%
, Adam, C\BPBI D.%
, Pelgrift, J\BPBI Y.%
, Lessac-Chenen, E\BPBI J.%
, Nelson, D\BPBI S.%
, Antreasian, P\BPBI G.%
\BDBL {}Lauretta, D\BPBI S.%
\end{APACrefauthors}%
\unskip\
\newblock
\APACrefYearMonthDay{in preparation}{}{}.
\newblock
{\BBOQ}\APACrefatitle {{Initial Orbit Determination and Event Reconstruction
  from Estimation of Particle Trajectories}} {{Initial Orbit Determination and
  Event Reconstruction from Estimation of Particle Trajectories}}.{\BBCQ}
\newblock
\APACjournalVolNumPages{JGR Planets}{}{}{}.
\PrintBackRefs{\CurrentBib}

\bibitem [\protect \citeauthoryear {%
Lucchesi%
}{%
Lucchesi%
}{%
{\protect \APACyear {2001}}%
}]{%
solrad}
\APACinsertmetastar {%
solrad}%
\begin{APACrefauthors}%
Lucchesi, D\BPBI M.%
\end{APACrefauthors}%
\unskip\
\newblock
\APACrefYearMonthDay{2001}{}{}.
\newblock
{\BBOQ}\APACrefatitle {{Reassessment of the Error Modelling of
  Non-Gravitational Perturbations on LAGEOS II and their Impact in the
  Lense-Thirring Determination. Part I}} {{Reassessment of the Error Modelling
  of Non-Gravitational Perturbations on LAGEOS II and their Impact in the
  Lense-Thirring Determination. Part I}}.{\BBCQ}
\newblock
\APACjournalVolNumPages{{Planetary and Space Science}}{49}{5}{}.
\PrintBackRefs{\CurrentBib}

\bibitem [\protect \citeauthoryear {%
Mahalanobis%
}{%
Mahalanobis%
}{%
{\protect \APACyear {1936}}%
}]{%
mahalanobis}
\APACinsertmetastar {%
mahalanobis}%
\begin{APACrefauthors}%
Mahalanobis, P\BPBI C.%
\end{APACrefauthors}%
\unskip\
\newblock
\APACrefYearMonthDay{1936}{}{}.
\newblock
{\BBOQ}\APACrefatitle {{On the Generalized Distance in Statistics}} {{On the
  Generalized Distance in Statistics}}.{\BBCQ}
\newblock
\APACjournalVolNumPages{Proceedings of the National Institute of Sciences
  (Calcutta)}{2}{}{}.
\PrintBackRefs{\CurrentBib}

\bibitem [\protect \citeauthoryear {%
McMillan%
, Scotti%
, Frecker%
, Gehrels%
\BCBL {}\ \BBA {} Perry%
}{%
McMillan%
\ \protect \BOthers {.}}{%
{\protect \APACyear {1986}}%
}]{%
modp2}
\APACinsertmetastar {%
modp2}%
\begin{APACrefauthors}%
McMillan, R.%
, Scotti, J.%
, Frecker, J.%
, Gehrels, T.%
\BCBL {}\ \BBA {} Perry, M\BPBI L.%
\end{APACrefauthors}%
\unskip\
\newblock
\APACrefYearMonthDay{1986}{}{}.
\newblock
{\BBOQ}\APACrefatitle {{Use of a Scanning CCD to Discriminate Asteroid Images
  Moving in a Field of Stars}} {{Use of a Scanning CCD to Discriminate Asteroid
  Images Moving in a Field of Stars}}.{\BBCQ}
\newblock
\BIn{} \APACrefbtitle {{Instrumentation in Astronomy VI}} {{Instrumentation in
  Astronomy VI}}\ (\BVOL~627, \BPGS\ 141--155).
\PrintBackRefs{\CurrentBib}

\bibitem [\protect \citeauthoryear {%
Mortari%
, Samaan%
, Bruccoleri%
\BCBL {}\ \BBA {} Junkins%
}{%
Mortari%
\ \protect \BOthers {.}}{%
{\protect \APACyear {2004}}%
}]{%
star2}
\APACinsertmetastar {%
star2}%
\begin{APACrefauthors}%
Mortari, D.%
, Samaan, M\BPBI A.%
, Bruccoleri, C.%
\BCBL {}\ \BBA {} Junkins, J\BPBI L.%
\end{APACrefauthors}%
\unskip\
\newblock
\APACrefYearMonthDay{2004}{}{}.
\newblock
{\BBOQ}\APACrefatitle {{The Pyramid Star Identification Technique}} {{The
  Pyramid Star Identification Technique}}.{\BBCQ}
\newblock
\APACjournalVolNumPages{Navigation}{51}{3}{171--183}.
\PrintBackRefs{\CurrentBib}

\bibitem [\protect \citeauthoryear {%
Padgett%
\ \BBA {} Kreutz-Delgado%
}{%
Padgett%
\ \BBA {} Kreutz-Delgado%
}{%
{\protect \APACyear {1997}}%
}]{%
star1}
\APACinsertmetastar {%
star1}%
\begin{APACrefauthors}%
Padgett, C.%
\BCBT {}\ \BBA {} Kreutz-Delgado, K.%
\end{APACrefauthors}%
\unskip\
\newblock
\APACrefYearMonthDay{1997}{}{}.
\newblock
{\BBOQ}\APACrefatitle {{A Grid Algorithm for Autonomous Star Identification}}
  {{A Grid Algorithm for Autonomous Star Identification}}.{\BBCQ}
\newblock
\APACjournalVolNumPages{IEEE Transactions on Aerospace and Electronic
  Systems}{33}{1}{202--213}.
\PrintBackRefs{\CurrentBib}

\bibitem [\protect \citeauthoryear {%
Pelgrift%
\ \protect \BOthers {.}}{%
Pelgrift%
\ \protect \BOthers {.}}{%
{\protect \APACyear {in preparation}}%
}]{%
pelgrift}
\APACinsertmetastar {%
pelgrift}%
\begin{APACrefauthors}%
Pelgrift, J\BPBI Y.%
, Lessac-Chenen, E\BPBI J.%
, Adam, C\BPBI D.%
, Leonard, J\BPBI M.%
, Nelson, D\BPBI S.%
, McCarthy, L.%
\BDBL {}Lauretta, D\BPBI S.%
\end{APACrefauthors}%
\unskip\
\newblock
\APACrefYearMonthDay{in preparation}{}{}.
\newblock
{\BBOQ}\APACrefatitle {{Reconstruction of Active Bennu Particle Events from
  Sparse Data}} {{Reconstruction of Active Bennu Particle Events from Sparse
  Data}}.{\BBCQ}
\newblock
\APACjournalVolNumPages{JGR Planets}{}{}{}.
\PrintBackRefs{\CurrentBib}

\bibitem [\protect \citeauthoryear {%
Pelgrift%
\ \protect \BOthers {.}}{%
Pelgrift%
\ \protect \BOthers {.}}{%
{\protect \APACyear {2018}}%
}]{%
camera_model}
\APACinsertmetastar {%
camera_model}%
\begin{APACrefauthors}%
Pelgrift, J\BPBI Y.%
, Sahr, E\BPBI M.%
, Nelson, D\BPBI S.%
, Jackman, C\BPBI D.%
, Benhacine, L.%
, Bos, B\BPBI J.%
\BDBL {}others%
\end{APACrefauthors}%
\unskip\
\newblock
\APACrefYearMonthDay{2018}{}{}.
\newblock
{\BBOQ}\APACrefatitle {{In-Flight Calibration of the OSIRIS-REx Optical
  Navigation Imagers}} {{In-Flight Calibration of the OSIRIS-REx Optical
  Navigation Imagers}}.{\BBCQ}
\newblock
\BIn{} \APACrefbtitle {{1st Annual RPI Workshop on Image-Based Modeling and
  Navigation for Space Applications}.} {{1st Annual RPI Workshop on Image-Based
  Modeling and Navigation for Space Applications}.}
\PrintBackRefs{\CurrentBib}

\bibitem [\protect \citeauthoryear {%
Scheeres%
\ \protect \BOthers {.}}{%
Scheeres%
\ \protect \BOthers {.}}{%
{\protect \APACyear {2019}}%
}]{%
bennu_gravity}
\APACinsertmetastar {%
bennu_gravity}%
\begin{APACrefauthors}%
Scheeres, D.%
, McMahon, J.%
, French, A.%
, Brack, D.%
, Chesley, S.%
, Farnocchia, D.%
\BDBL {}others%
\end{APACrefauthors}%
\unskip\
\newblock
\APACrefYearMonthDay{2019}{}{}.
\newblock
{\BBOQ}\APACrefatitle {{The Dynamic Geophysical Environment of (101955) Bennu
  Based on OSIRIS-REx Measurements}} {{The Dynamic Geophysical Environment of
  (101955) Bennu Based on OSIRIS-REx Measurements}}.{\BBCQ}
\newblock
\APACjournalVolNumPages{Nature Astronomy}{3}{4}{352}.
\PrintBackRefs{\CurrentBib}

\bibitem [\protect \citeauthoryear {%
Spratling%
\ \BBA {} Mortari%
}{%
Spratling%
\ \BBA {} Mortari%
}{%
{\protect \APACyear {2009}}%
}]{%
star3}
\APACinsertmetastar {%
star3}%
\begin{APACrefauthors}%
Spratling, B.%
\BCBT {}\ \BBA {} Mortari, D.%
\end{APACrefauthors}%
\unskip\
\newblock
\APACrefYearMonthDay{2009}{}{}.
\newblock
{\BBOQ}\APACrefatitle {{A Survey on Star Identification Algorithms}} {{A Survey
  on Star Identification Algorithms}}.{\BBCQ}
\newblock
\APACjournalVolNumPages{Algorithms}{2}{1}{93--107}.
\PrintBackRefs{\CurrentBib}

\bibitem [\protect \citeauthoryear {%
Wright%
, Liounis%
\BCBL {}\ \BBA {} Ashman%
}{%
Wright%
\ \protect \BOthers {.}}{%
{\protect \APACyear {2018}}%
}]{%
giant}
\APACinsertmetastar {%
giant}%
\begin{APACrefauthors}%
Wright, C\BPBI A.%
, Liounis, A\BPBI J.%
\BCBL {}\ \BBA {} Ashman, B\BPBI W.%
\end{APACrefauthors}%
\unskip\
\newblock
\APACrefYearMonthDay{2018}{}{}.
\newblock
{\BBOQ}\APACrefatitle {{Optical Navigation Algorithm Performance}} {{Optical
  Navigation Algorithm Performance}}.{\BBCQ}
\newblock
\BIn{} \APACrefbtitle {{1st Annual RPI Workshop on Image-Based Modeling and
  Navigation for Space Applications}.} {{1st Annual RPI Workshop on Image-Based
  Modeling and Navigation for Space Applications}.}
\PrintBackRefs{\CurrentBib}

\bibitem [\protect \citeauthoryear {%
Zacharias%
\ \protect \BOthers {.}}{%
Zacharias%
\ \protect \BOthers {.}}{%
{\protect \APACyear {2013}}%
}]{%
ucac}
\APACinsertmetastar {%
ucac}%
\begin{APACrefauthors}%
Zacharias, N.%
, Finch, C.%
, Girard, T.%
, Henden, A.%
, Bartlett, J.%
, Monet, D.%
\BCBL {}\ \BBA {} Zacharias, M.%
\end{APACrefauthors}%
\unskip\
\newblock
\APACrefYearMonthDay{2013}{}{}.
\newblock
{\BBOQ}\APACrefatitle {{The Fourth US Naval Observatory CCD Astrograph Catalog
  (UCAC4)}} {{The Fourth US Naval Observatory CCD Astrograph Catalog
  (UCAC4)}}.{\BBCQ}
\newblock
\APACjournalVolNumPages{{The Astronomical Journal}}{145}{2}{44}.
\PrintBackRefs{\CurrentBib}

\end{thebibliography}

\end{document}